# Hydrogen-assisted layer-by-layer growth and robust nontrivial topology of stanene films on Bi(111)

Liying Zhang[1, 2, 3], Leiqiang Li[3], Chenxiao Zhao[4], Shunfang Li[2], Jinfeng Jia[4, 5, 6], Zhenyu Zhang[3], Yu Jia[1, 2, *], and Ping Cui[3, *]

[1] *Key Laboratory for Special Functional Materials of Ministry of Education, Collaborative Innovation Center of Nano Functional Materials and Applications, School 817046of Materials Science and Engineering, Henan University, Kaifeng, 475004, China*

[2] *International Laboratory for Quantum Functional Materials of Henan, and School of Physics and Microelectronics, Zhengzhou University, Zhengzhou 450001, China*

[3] *International Center for Quantum Design of Functional Materials (ICQD), Hefei National Laboratory for Physical Sciences at Microscale, and CAS Center for Excellence in Quantum Information and Quantum Physics, University of Science and Technology of China, Hefei, Anhui 230026, China*

[4] *Key Laboratory of Artificial Structures and Quantum Control (Ministry of Education), Shenyang National Laboratory for Materials Science, School of Physics and Astronomy, Shanghai Jiao Tong University, Shanghai 200240, China*

[5] *Tsung-Dao Lee Institute, Shanghai Jiao Tong University, Shanghai 200240, China*

[6] *CAS Center for Excellence in Topological Quantum Computation, University of Chinese Academy of Sciences, Beijing 100190, China*

\*Corresponding Authors: jiayu@henu.edu.cn; cuipg@ustc.edu.cn

## Abstract

Ever since the first successful synthesis of stanene via epitaxial growth, numerous efforts have been devoted to improving its overall quality and exploring its topological and other exotic properties under different growth conditions. Here, using first-principles approaches, we reveal the atomistic growth mechanisms and robust topological properties of few-layer stanene on Bi(111). We first show that monolayer stanene grown on Bi(111) follows a highly desirable nucleation-and-growth mechanism, characterized by attractive interaction of the Sn adatoms. More importantly, we reveal that surface passivation by the residual hydrogen is essential in achieving layer-by-layer growth of high-quality few-layer stanene, with the hydrogen functioning as a surfactant in stabilizing the growing films. Furthermore, we investigate systematically the dependence of the topological properties of stanene on the film thickness, hydrogen passivation, and substrate, and obtain robust quantum spin Hall effects under diverse physical conditions. The robustness of the nontrivial topology is attributable to the strong spin-orbit coupling of the Bi substrate. The atomistic growth mechanisms and nontrivial topology of stanene as presented here are also discussed in connection with recent experimental findings.

# I. INTRODUCTION

Over the last decades, topological materials have been widely explored as attractive new classes of quantum matter. Such materials are characterized by their nontrivial band topologies, and are commonly manifested by the existence of robust topological boundary states following the bulk-boundary correspondence [1,2]. Specific examples include the quantum spin Hall (QSH) insulators [3-6], three-dimensional topological insulators [7-11], topological crystalline insulators [12,13], and the recently reported higher-order topological insulators [14-16]. Among these topological classes, a QSH insulator, as a representative topological insulator in the two-dimensional (2D) regime, has attracted extensive research attention due to its potential applications in dissipationless electronics [1,2] and quantum computing [17,18]. A variety of atomically thin 2D materials have been explored theoretically and experimentally for potential realization of the QSH effects [3,19-28], yet to date, strong evidences of their existence in such ultrathin systems have only been reported in very few systems, such as monolayered $WTe_2$ [26-28] and bilayered Bi [29]. In these studies, the standing challenges on the experimental side are tied to structural or chemical instabilities of the grown samples, such as defects and unexpected surface decoration [22,30-32], making it highly desirable to fabricate high-quality samples for subsequent characterizations on their topological properties.

Among the candidate QSH systems, the proposal of stanene as a large-gap QSH insulator [21] has stimulated extensive experimental efforts on synthesizing stanene films under devise conditions, subsequently characterizing their emergent properties. It was first successfully fabricated on the $Bi_2Te_3(111)$ substrate by molecular beam epitaxy [24], and later also on other substrates, such as Sb(111) [33] and InSb(111) [34]. Moreover, a major advance in realizing nontrivial band topology has been achieved when growing ultra-flat stanene films on the Cu(111) surface [35]. Very recently, few-layer stanene films grown on PbTe(111) has been revealed to exhibit layer-dependent superconductivity [36] with type-II Ising pairing [37]. These exciting developments amply demonstrate that stanene can provide a rich platform for exploring exotic topological physics and quantum device applications in the 2D regime. Such potential advances, in turn, critically depend on discoveries of elegant kinetic growth pathways for fabricating high-quality stanene samples.

In an earlier study, we reported that a stanene monolayer obeys different atomistic mechanisms when grown on different $Bi_2Te_3(111)$-based substrates [38]. In particular, the Bi(111)-bilayer covered $Bi_2Te_3(111)$ substrate was predicted to strongly favor the growth of single crystalline stanene [38]. This prediction has been largely supported by the latest experimental demonstration of high-quality few-layer stanene grown on the Bi(111) substrate with robust edge states, as reported elsewhere [39]. In the present companying study, we use first-principles approaches to investigate the atomistic growth mechanisms and related topological properties of few-layer stanene on the Bi(111) substrate, with substantial and distinctly new findings. First, we find that a stanene monolayer on Bi(111) also follows a highly desirable nucleation-and-growth mechanism, characterized by fast diffusion and attractive interaction of the Sn adatoms. Next, we reveal that hydrogen passivation on the growth front of the stanene films is essential in achieving layer-by-layer growth in the few-layer regime, with the hydrogen functioning as a surfactant in stabilizing the growing films. Such surfactant action is vital in acquiring the high-quality stanene samples reported experimentally [39]. Furthermore, we show that the stanene films on Bi(111) possess a topological nontrivial invariant with $Z_2 = 1$ against the layer thickness. Such a robustness of the nontrivial topology is not due to the surface hydrogenation of the stanene films, but rather, is likely to be derived from the proximity effect of the Bi(111) substrate with inherently strong spin-orbit coupling (SOC). Both the novel atomistic growth mechanisms and robust nontrivial topology of stanene films predicted here are also discussed in connection with the recent experimental findings.

The paper is organized as follows. In Sec. II, we briefly describe the computational methods, the definitions of adsorption energy and formation energy, and the criteria of the film stability as measured by the second-order difference of the formation energy. In Sec. III, we reveal the growth mechanisms of monolayer and few-layer stanene on Bi(111) in the presence of hydrogen acting as a surfactant, and further demonstrate their robust nontrivial topological properties against the layer thickness. In Sec. IV, we briefly discuss these results in connection with the recent experimental findings. Finally, we conclude in Sec. V.

## II. COMPUTATIONAL METHODS

Our density function theory (DFT) calculations were carried out using the Vienna *ab initio* simulation package (VASP) [40], where the projector augmented wave (PAW) method [41,42] was adopted, and the generalized gradient approximation (GGA) in the framework of Perdew-Burke-Ernzerhof (PBE) [43] was chosen for the exchange-correlation functional. The van der Waals (vdW) correction was treated using the DFT-D3 method [44] and included in all the calculations. The single-crystal Bi has a bilayered structure, with an ABC stacking sequence along the (111) direction [19]. The lattice constants of a hexagonal unit cell of bulk Bi are $a = b = 4.55$ Å chosen from our experimental data [39], and the interlayer spacing is $c = 12.102$ Å obtained by optimization using the DFT-D3 method. The value of $c$ is in good agreement with the experimental value (11.797 Å) [45], indicating that the DFT-D3 method can properly describe the vdW interactions for this system. The lattice mismatch of the Bi(111) substrate and stanene ($a = b = 4.68$ Å [21]) is 2.77%. The Bi(111) substrate was modeled by a six-bilayer slab for structural optimization and energy calculations, and a three-bilayer slab was used for calculations of electronic and topological properties, both with a vacuum layer thicker than 15 Å to ensure the decoupling between neighboring slabs. A 7×7×1 Monkhorst-Pack $k$-point meshes were used for the 1×1 surface unit cell, and a 2×2×1 sampling for the (4×4) or (5×5) surface supercell. The energy cutoff was set to 400 eV. During the structural optimization, the atoms in the bottom-most four bilayers were fixed to their respective bulk positions, while all other atoms were fully allowed to relax until the corresponding forces were smaller than 0.01 eV/Å. The SOC effects were considered unless otherwise specified.

To investigate the topological properties of these stanene films with and without the Bi(111) substrate, we calculate the $Z_2$ invariant via the Wannier charge centers [46] and construct the edge Green's function of the semi-infinite lattice model [47] from the maximally localized Wannier functions (MLWFs) [48] as implemented in the WannierTools package [49]. The Wannier process requires the orbital projections onto not only stanene films but also the whole Bi substrate, drastically increasing the dimensionality of the projection basis and leading to demanding technical difficulty. To deal with this issue, we downfold the projection basis to a subspace spanned by the stanene films and topmost three bilayers of the Bi(111) substrate. Therefore, a three-bilayer slab of the Bi(111) substrate was used to calculate the electronic structure and analyze the corresponding topological properties.

The adsorption energy ($E_{ads}$) per Sn adatom is defined by $E_{ads} = -(E_{tot} - E_{sub} - N_{Sn} \times E_{Sn})/N_{Sn}$, where $E_{tot}$, $E_{sub}$, and $E_{Sn}$ are the total energies of the combined system, the substrate, and one Sn atom in gas phase, respectively, and $N_{Sn}$ is the total number of Sn adatoms. The formation energy ($E_f$) of few-layer stanene on Bi(111) is defined by the formula $E_f = -(E_{tot} - E_{sub} - N_H \times E_H - N \times E_{stanene})/N$, where $E_H$ and $E_{stanene}$ are the total energies of one H atom in gas phase and a freestanding stanene monolayer containing the same number of Sn atoms as that in each layer of stanene films, respectively, $N_H$ is the number of passivated H atoms, and $N$ is the layer thickness of the stanene film. In order to examine the stability of the stanene films, we also define the second-order difference of $E_f$, by $E''_f(N) = -2 \times E_f(N) + E_f(N-1) + E_f(N+1)$. If $E''_f(N) \geq 0$, the film of $N$ stanene layers is stable, otherwise the film is unstable [50].

## III. RESULTS

### A. Energetics and kinetics of Sn adatoms at the initial stages of stanene growth on Bi(111)

To investigate the growth mechanisms of stanene, we first consider the energetics and kinetics of isolated Sn adatoms on Bi(111). The Sn atoms were initially placed on three high-symmetry adsorption sites (the fcc hollow, hcp hollow, and top sites) as well as four asymmetric adsorption sites (namely, the bridge site and the middle points of neighboring high-symmetry sites, see Fig. S1(a) of the Supplemental Material [51]). On each of the initial adsorption sites, the Sn atom was placed about 3 Å above the top-most Bi surface. Upon structural relaxation, the Sn adatoms prefer to stay on the three high-symmetry sites (hcp, fcc and top), and the corresponding adsorption energies ($E_{ads}$) are given as 2.49, 2.14, and 1.39 eV per Sn, respectively (see Fig. 1(a)). Here, the locally stable top configuration is essentially also an unfavorable state, given its much higher adsorption energy. Therefore, only the stable hcp and metastable fcc sites are considered in subsequent studies. The Sn-Bi bond length and vertical distance between the Sn atom and a neighboring Bi atom are 2.87 and 2.36 Å in the most stable hcp configuration. To grow stanene, the Sn adatoms should be able to diffuse rapidly on the Bi(111) surface. The calculated activation energy barrier against a Sn adatom hopping from the hcp to its neighboring fcc hollow site is 0.44 eV (see Fig. 1(b)), and the barrier against the reverse process is even lower (0.10 eV), indicating overall fast diffusion of the Sn adatoms on

Bi(111) at typical growth temperatures. These results are advantageous to the potential growth of quality stanene samples.

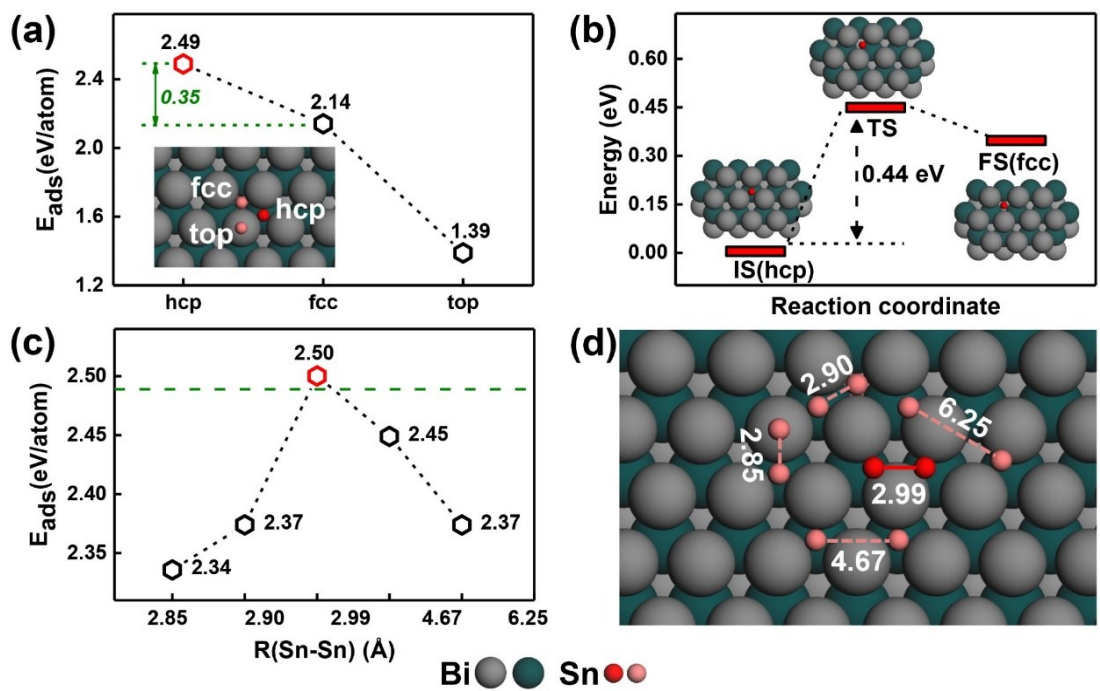

FIG. 1. (a) Adsorption energies of isolated Sn adatoms on Bi(111), with the high-symmetry adsorption sites shown in the insert. (b) Minimum energy path for Sn diffusion on Bi(111), with the initial state (IS), transition state (TS), and final state (FS) specified. (c) Average adsorption energies of two Sn adatoms as a function of the Sn-Sn separation $R_{Sn-Sn}$. The dashed line indicate the adsorption energy of an isolated Sn adatom. (d) Representative adsorption configurations of two Sn adatoms on Bi(111). In (a) and (d), the gray and green spheres represent the upper and lower Bi atoms in each Bi bilayer, respectively.

In order to reveal the atomistic mechanism at the initial stages of stanene growth, we further consider the interaction of two Sn adatoms on Bi(111) surface. Possible initial configurations considered are shown in Fig. S1(b), where one Sn adatom is placed on the most stable hcp hollow site, while the other is placed on possible nearby high-symmetry sites within the Sn-Sn distance ($R_{Sn-Sn}$) range of less than 7.9 Å. Such initial Sn-Sn distances cover from the typical bond length of a Sn dimer in gas phase [52] to much larger. The $E_{ads}$ and their representative adsorption configurations upon structural relaxation are presented in Figs. 1(c) and 1(d). We observe that the $R_{Sn\text{-}Sn}$ of the most stable structure is 2.99 Å, corresponding to the case at which the Sn adatoms largely occupy neighboring bridge sites, forming a well-defined Sn-Sn bond, and the vertical distance between the Sn atoms and the substrate is 1.72 Å. The adsorption energy of the most stable configuration of the two Sn atoms is 2.50 eV per Sn, higher than that of the other configurations with smaller or larger $R_{Sn\text{-}Sn}$. Given this adsorption-energy behavior, two isolated Sn adatoms can readily nucleate into a dimer when they get close enough on the surface due to the existence of an effective attractive pairwise interaction. These Sn adsorption, diffusion, and dimerization behaviors collectively characterize the growth mode of stanene on Bi(111) to follow the standard nucleation-and-growth mechanism, similar to stanene growth on the Bi(111)-bilayer covered $Bi_2Te_3$ substrate predicted previously in the monolayer growth regime [38].

**B. Hydrogen functioning as a surfactant in achieving layer-by-layer growth of stanene on Bi(111)**

The recent experiment [39] and other previous studies [53,54] have shown that hydrogen is ubiquitous and difficult to remove during stanene growth. Given this fact, the hydrogen effect is indispensable in investigation of the growth mechanisms of few-layer stanene on Bi(111) and their corresponding topological properties. In this subsection, we focus on the growth mechanisms by considering the adsorption configurations and energetics of few-layer stanene on Bi(111) with the layer thickness $N$ = 1, 2, …, 6. In particular, both cases of the clean and hydrogen-passivated stanene films on Bi(111) are comparatively studied, with the important finding that the unavoidable and seemingly undesirable presence of hydrogen turns out to be crucial for acquiring high-quality growth.

Given the relatively small lattice mismatch between stanene and the Bi substrate (2.77 %), a 1×1 surface unit cell was used for the combined system. Six possible initial high-symmetry packing configurations were considered, labeled by the locations of the lower-upper Sn atoms in the bottom stanene overlayer as fcc-hcp, fcc-top, hcp-fcc, hcp-top, top-fcc, and top-hcp, as demonstrated in a 4-layer stanene on Bi(111) in Fig. 2. The stacking configuration of the stanene layers follows the structure of α-Sn, as verified in the recent experimental study [39]. After structural optimization, we establish fcc-hcp as the most stable configuration for all the considered layer thicknesses without or with the hydrogen passivation, with the corresponding formation energies ($E_f$) shown in Figs. 3(a) and 3(b). Compared with the clean surface case, the hydrogen atoms saturate the dangling bonds of the Sn atoms in the top-most stanene overlayer, thereby significantly enhancing the overall formation energies and their corresponding stabilities. We also observe that, as the thickness of the stanene films increases, the energy differences between different adsorption configurations become smaller and smaller, again without or with the hydrogen passivation. Qualitatively, such a trend is manifestation of the weakening substrate effect on the film stability.

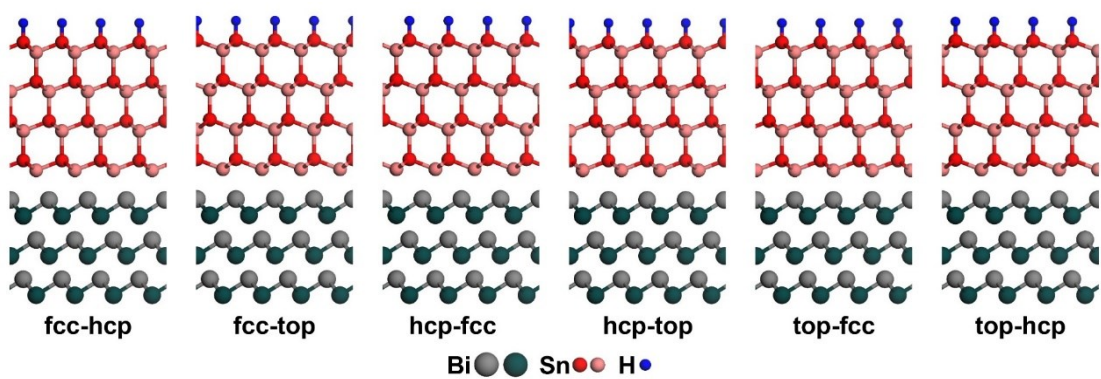

FIG. 2. Side views of six high-symmetry packing configurations of a 4-layer stanene film on Bi(111) with hydrogen passivation.

At a closer or quantitative level, the energy difference between the most and second-most stable configurations is 19 meV per unit cell of a stanene monolayer for the clean surface case, while this value is doubled to 37 meV per unit cell with hydrogen passivation. Such an enhanced energy difference is beneficial to the suppression of the orientational grain boundaries during the stanene growth [38,55,56], indicating that the hydrogen atoms will help to improve the quality of the stanene monolayer on Bi(111), as confirmed experimentally [39]. Here we should also note that, for multilayer stanene growth, their preferred packing configurations are largely determined by the growth of the bottom layer, as energetically selected by the α-Sn structure [39].

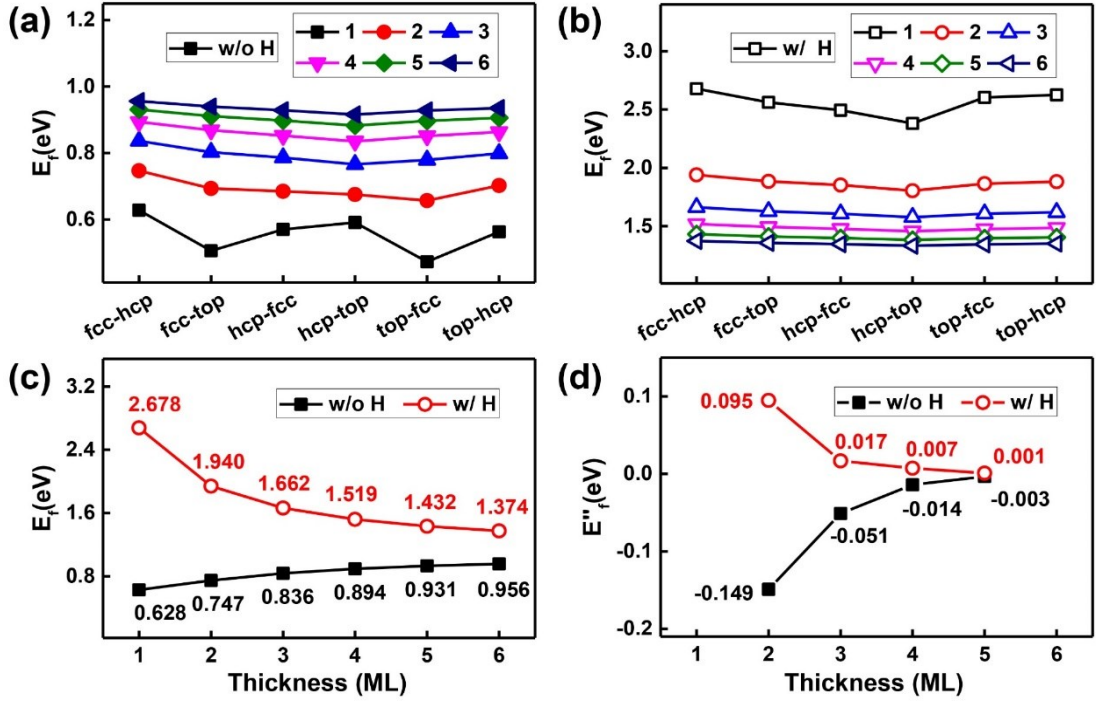

FIG. 3. Formation energies ($E_f$) of stanene at different layer thicknesses ($N = 1, 2, …, 6$) on Bi(111) (a) without and (b) with H passivation of the top layer, including six high-symmetry packing configurations. (c) Formation energies of the most stable configuration for each stanene thickness without (black) and with (red) H passivation. (d) The corresponding second-order difference ($E''_f$) as a function of the layer thickness of the stanene films.

To further examine the relative stabilities of the few-layer stanene films, we contrast the formation energies of the most stable configurations for each layer thickness without and with hydrogen passivation in Fig. 3(c), with the corresponding second-order differences ($E''_f$) shown in Fig. 3(d). We observe that the formation energy increases with increasing layer thickness for the clean case, but exhibits a decreasing trend upon hydrogen passivation. To appreciate the significance of these contrasting trends, we note that the experimentally established growth mode is selected by the interplay of the energetics and kinetics under nonequilibrium growth conditions [57]. In particular, the second-order difference of the formation energies shown in Fig. 3(d) measures the stability of the growing films [50,58]. Here, $E''_f(N) < 0$ for $N = 2, 3, 4, 5$ without hydrogen passivation, indicating that the stanene films are unstable; in contrast, $E''_f(N) > 0$ for $N = 2, 3, 4, 5$ with hydrogen passivation, indicating that the films are stable.

Therefore, the hydrogen passivation of the growth front is essential in achieving layer-by-layer growth of the high-quality stanene films, with the hydrogen functioning as a surfactant [53,57].

## C. Robust topological properties of stanene films on Bi(111)

As mentioned above, the potential existence of QSH states in freestanding stanene has been proposed, including the effects of hydrogen passivation and layer thickness [21]. Here, it is naturally desirable to investigate the topological properties of stanene films on Bi(111) as well, not only because the Bi films have been demonstrated to host nontrivial topology [19,21,59-63], but also because of the very recent observations of robust edge states in these systems [39].

To investigate the topological properties of stanene films on Bi(111), we first examine their electronic structures by considering the representative layer thicknesses from 1 to 6 layers. The band structures of the 1-4 layer stanene films on Bi(111) without H passivation within the PBE+SOC scheme are shown in Figs. 4(a)-4(d), exhibiting overall similar features. Around the K point, the bands are mainly contributed by the $p_z$ orbitals of stanene, which can be attributed to the existence of the dangling bonds on the surface of the stanene films. Around the Γ point, the $p_x$ and $p_y$ orbitals of stanene exhibit heavy hybridization, and the spectral weights of the $p_y$ orbitals clearly show a layer-dependence feature. Compared with the PBE results shown in Fig. S3, the degeneracy of the bands has been lifted within the whole surface Brillouin zone except at the K points protected by the time-reversal symmetry (i.e., the Γ and M points). Such Rashba splittings result from the strong SOC effect of the heavy element Sn together with the inversion symmetry breaking due to the existence of the Bi(111) substrate. More importantly, for the 1-layer system, the SOC has inverted the $p_y$ orbital contribution to the upper neighboring band at ~0.2 eV around the Γ point, while for the 2-layer system, the SOC plays a role in opening a gap at the Γ point. The 3-layer and 4-layer systems show the similar band inversions as the 1-layer system with the inclusion of the SOC, but at higher energy levels. Such SOC-induced band inversions and gap openings may lead to topological phase transitions, as demonstrated later.

When the topmost surface of stanene is passivated by hydrogen, the band structures of the 1-6 layer stanene films on Bi(111) largely exhibit similar features as elaborated in the absence of the H passivation, as shown in Figs. 4(e)-4(h) and Fig. S3 for the cases of 1-4 layers. The difference is that many of the $p_z$ orbital components around the K point have been tuned deeper into the low energy levels far away from the Fermi level, due to the saturation of the surface dangling bonds by hydrogen. Moreover, the SOC again plays an important role in inverting the $p_y$ orbital contributions to the higher energy levels for the 1-layer and 2-layer systems, while opening band gaps at the Γ points for the 3-layer and 4-layer systems.

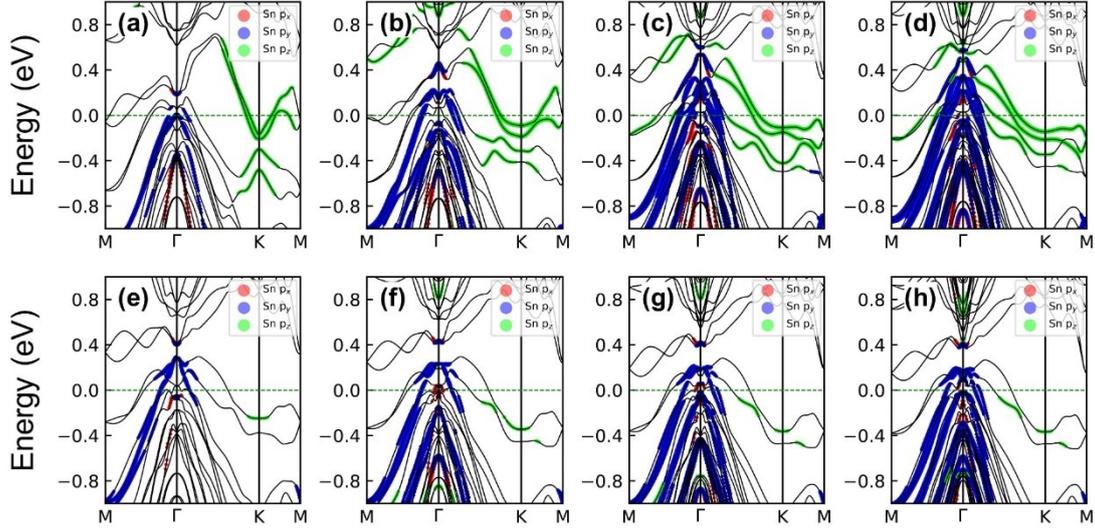

FIG. 4. Band structures of the (a) 1-layer, (b) 2-layer, (c) 3-layer, and (d) 4-layer stanene on Bi(111) without H passivation, obtained within the PBE+SOC scheme. The corresponding results with H passivation are plotted in (e), (f), (g), and (h). Here, the sizes of the red, blue, and green circles denote the spectral weights contributed by the $p_x$, $p_y$, and $p_z$ orbitals of the Sn atoms, respectively. The Fermi levels are all set at zero.

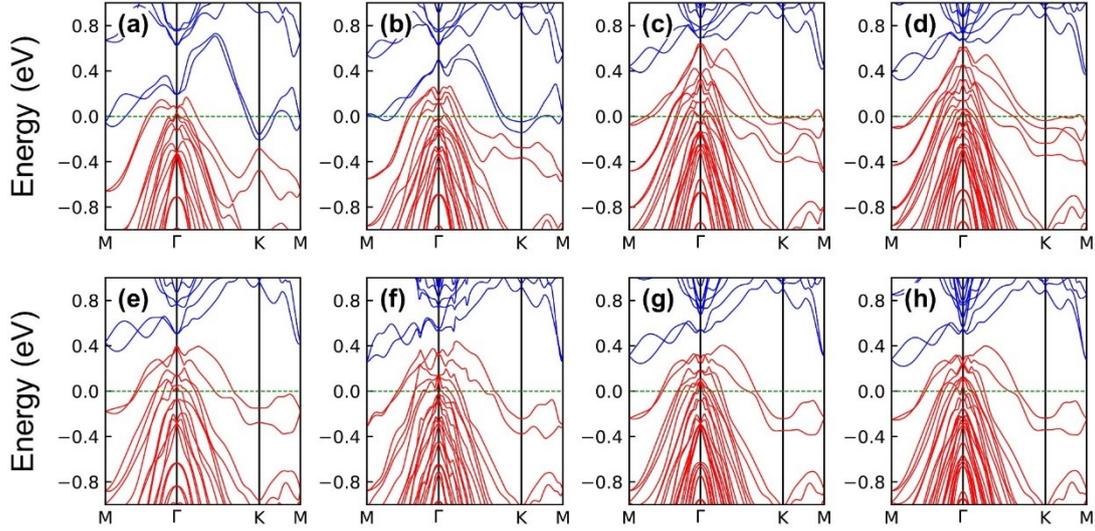

FIG. 5. Band structures of the (a) 1-layer, (b) 2-layer, (c) 3-layer, and (d) 4-layer stanene on Bi(111) without H passivation, obtained using the PBE+SOC scheme as fitted by the MLWFs. The corresponding results with H passivation are plotted in (e), (f), (g), and (h). Here, all the red (blue) bands are treated as the occupied (unoccupied) Wannier bands for the $Z_2$ invariant calculations. The Fermi levels are all set at zero.

Given the above analyses on the electronic structures, we now determine the topological properties of stanene films on Bi(111) with or without hydrogen passivation by evaluating the topological invariant $Z_2$. During the Wannier process, we project the Bloch wavefunctions onto the $p$ orbitals of both the stanene films and Bi(111) substrate, and obtain the band structures by fitting with the MLWFs as shown in Fig. 5. It is clear that the selected $p$ orbitals can fit the PBE+SOC results quite well. Although there are no global gaps in these systems, we can still introduce a "curved chemical potential" by explicitly separating the bands into the "occupied" and "unoccupied" states, as highlighted by the red and blue solid lines in Fig. 5. It is therefore convenient to calculate the related $Z_2$ invariants by the Wannier charge centers [46]. We find that each of the 1-6 layer stanine films on Bi(111) hosts a topologically nontrivial phase with $Z_2 = 1$, characterizing the existence of the QSH states, irrespective of the hydrogen passivation. As another manifestation of the nontrivial topology in the band structures, we further obtain the corresponding edge states of the 1-4 layer systems using semi-infinite lattice models [47] as shown in Fig. 6, exhibiting the Dirac nature of the band structures.

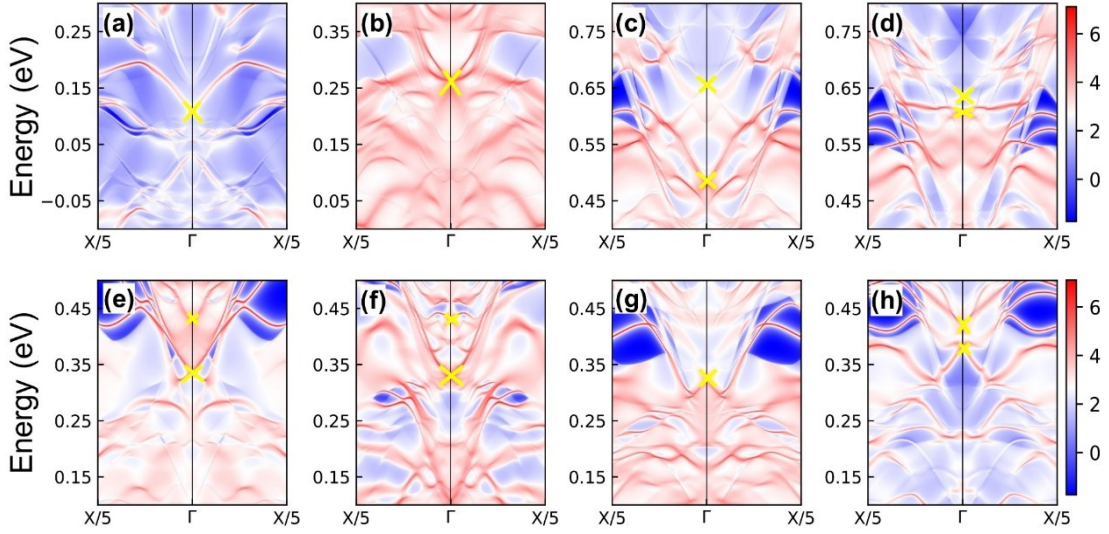

FIG. 6. Edge states of the (a) 1-layer, (b) 2-layer, (c) 3-layer, and (d) 4-layer stanene on Bi(111) without H passivation, obtained using the semi-infinite slabs along the (100) direction. The corresponding edge states of the systems with H passivation are plotted in (e), (f), (g), and (h). Here, the Dirac points are marked by the yellow crosses. The Fermi levels are all set at zero.

To reveal the physical origin of the robust nontrivial topology of these stanene films on Bi(111), we identify the topological properties of freestanding 1-6 layer stanene in a comparative study, but with the structures constrained to be the same as determined on the Bi(111) substrate. The band structures of the 1-4 layer constrained stanene films without or with hydrogen passivation within the PBE+SOC scheme are shown in Fig. 7. Using the same "curved chemical potential" approach as for the stanene films on Bi(111), we can again calculate the $Z_2$ invariants to be 1, 0, 1, 0, 1 and 1 in the absence of hydrogen passivation and 0, 0, 1, 1, 1, and 1 in the presence of hydrogen passivation for 1-6 layer films. The corresponding band structures fitted by the MLWFs are shown in Fig. S4, and the edge states are plotted in Fig. S5. We also note that these results are somewhat different from Ref. [64] in certain detailed aspects, caused by the adoption of the constrained geometries in the present work.

Based on the above results, we can see that the trivial and nontrivial topological properties of the freestanding stanene films can oscillate or alter without or with the hydrogen passivation. Previous studies have also shown that stacking of the QSH insulator layers would alter the topological invariant and result in an odd-even oscillation of the $Z_2$ number [59,65]. Nevertheless, such layer dependences of the topology are absent in the stanene films on Bi(111). This observation is striking, and can be qualitatively attributed to the presence of the Bi substrate that possesses inherently strong SOC and is more capable of promoting the nontrivial topology in the few-layer stanene via proximity effects [19,66].

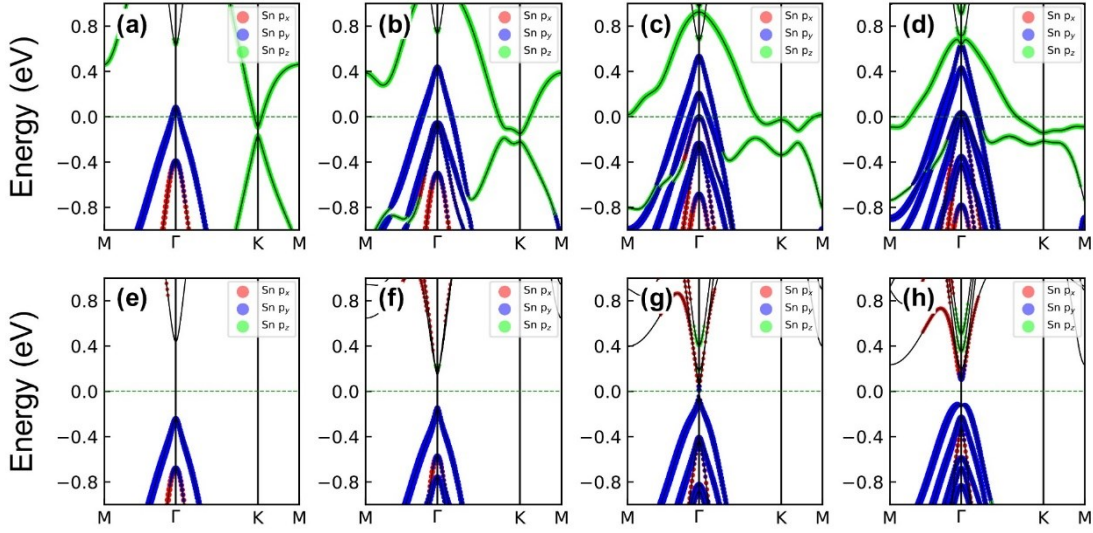

FIG. 7. Band structures of the (a) 1-layer, (b) 2-layer, (c) 3-layer, and (d) 4-layer freestanding stanene without H passivation, obtained within the PBE+SOC scheme. The corresponding results with H passivation are plotted in (e), (f), (g), and (h). Here, the sizes of the red, blue, and green circles denote the spectral weights contributed by the $p_x$, $p_y$, and $p_z$ orbitals of the Sn atoms, respectively. The Fermi levels are all set at zero.

## IV. DISCUSSION

As mentioned earlier, our previous prediction of improved growth of a stanene monolayer on the Bi(111)-bilayer covered $Bi_2Te_3$(111) has motivated the very recent successful demonstration of high-quality stanene films on the Bi(111) substrate [39]. Specifically, compared with the first successful synthesis of stanene on $Bi_2Te_3$(111) [24], the overall quality of the stanene films has been significantly improved both in the monolayer and few-layer regimes. As a more encouraging observation on the agreement of theory and experiment, an extra annealing process has to be applied after growing each stanene layer in the actual experiment to obtain higher-quality few-layer stanene, which in essence is to allow the topmost stanene layer to be fully passivated by the residual hydrogen atoms in the chamber. Such a crucial step is a compelling demonstration of the hydrogen effects as a surfactant in achieving layer-by-layer growth of high-quality stanene films.

The success in achieving high-quality stanene films in the few-layer regime also offers unprecedented opportunities to reveal their topological properties, as demonstrated by the recent experimental observations of the robust edge states in the 1-5 layer stanene films grown on Bi(111) [39]. Specifically, the *in-situ* scanning tunneling microscopy/spectroscopy measurements combined with our present $Z_2$ calculations show clear evidences of the topological edge states in all the 1-5 layer films. Such a finding is in stark contrast with the prevailing expectation that stacking of QSH insulator layers would alter the topological invariant and result in an odd-even oscillation of the $Z_2$ number [59,65]. As discussed earlier, the robustness of the nontrivial topology of the stanene overlayers is likely to originate from the strong SOC effects of the Bi(111) substrate, and such an agreement between theory and experiment further enhances the validity of the central findings of the present study.

As a prospective note, the capability of such epitaxial growth of high-quality stanene films should also offer new opportunities to potentially realize topological superconductivity in 2D systems, especially given the recent experimental demonstrations of superconductivity in few-layer stanene [36,37,39]. To date, the coexistence of nontrivial topological edge states and superconductivity in atomically thin systems has been experimentally reported only in separate measurements. In WTe$_2$, topological edge states were first reported [26,27], and superconductivity was subsequently achieved upon external electrical gating [67,68] or proximity effects associated with an additional *s*-wave superconductor [69]. In Bi(111) films, the observation of superconducting pairing also required the participation of an additional *s*-wave superconductor [29,70] Strikingly, the few-layer stanene samples inherently exhibit both topological edge states and superconductivity, rendering the systems to serve as a promising platform for realization of 1D topological superconductivity in a single-element material.

## V. CONCLUSIONS

In summary, we have systematically investigated the atomistic growth mechanisms and topological properties of stanene films on Bi(111), focusing on the effects of the film thickness and hydrogen passivation. We found that single-layer stanene grown on Bi(111) follows a nucleation-and-growth mode, characterized by attractive interaction between the Sn adatoms, while a hydrogen-assisted layer-by-layer growth mechanism has been revealed for few-layer stanene, with the hydrogen functioning as a surfactant in stabilizing the growing films. Furthermore, the quantum spin Hall states with the topological invariant $Z_2 = 1$ have been identified for stanene on Bi(111) regardless of the layer thickness or hydrogen passivation. Such a robustness is attributed to the proximity effects of the Bi substrate that possesses inherently strong spin-orbit coupling. The central findings as presented here are consistent with the recent experimental observations in the growth and topological properties of few-layer stanene on Bi(111) [39]. The present study provides the atomistic basis for realization of ultrathin stanene films that may harbor one-dimensional topological superconductivity in much simpler (namely, single-element) materials.

ACKNOWLEDGMENTS

This work was supported by the National Key R&D Program of China (Grant Nos. 2017YFA0303500, 2019YFA0308600, 2020YFA0309000, 2016YFA0301003, and 2016YFA0300403), the National Natural Science Foundation of China (Grant Nos. 11634011, 11722435, 11974323, 12074099, 11774078, 12074345, 11521404, 11634009, 92065201, and 11861161003), the Anhui Initiative in Quantum Information Technologies (Grant No. AHY170000), and the Strategic Priority Research Program of Chinese Academy of Sciences (Grant Nos. XDB30000000 and XDB28000000).